\documentclass[aps,ams,prl,twocolumn,groupedaddress,showpacs]{revtex4}

\usepackage{graphicx}

\begin{document}

\title{Statistics of speckle patterns}

\author{Oded Agam}
\altaffiliation[Permanent address:]{ The Racah Institute of Physics,
The Hebrew University, Jerusalem, 91904, Israel}

\affiliation{Department of Physics, University of Washington,
  Seattle, Washington, 98195-1560, USA}

\author{A. V. Andreev}

\affiliation{Department of Physics, University of Washington,
  Seattle, Washington, 98195-1560, USA}

\author{B. Spivak}

\affiliation{Department of Physics, University of Washington,
  Seattle, Washington, 98195-1560, USA}


\begin{abstract}
We develop a general method for calculating statistical properties
of the speckle pattern of coherent waves propagating in disordered
media. In some aspects this method is similar to the
Boltzmann-Langevin approach for the calculation of classical
fluctuations. We apply the method to the case  where the incident
wave experiences many small angle scattering events during
propagation, but the total angle change remains small. In many
aspects our results for this case are  different from results
previously known in the literature. The correlation function of the
wave intensity at two points separated by a distance $r$, has a long
range character. It decays as a power of $r$ and changes sign. We
also consider sensitivities of the speckles to changes of external
parameters, such as the wave frequency and the incidence angle.
\end{abstract}


\maketitle

In this article we consider statistical properties of waves
propagating trough a disordered medium and described by the
stationary (scalar) wave equation,
\begin{equation}
 \left(\nabla^{2}+k^2n^2({\bf r})\right)\psi({\bf r})=0,
\label{eq:wave}
\end{equation}
where $n({\bf r})$ is the index of refraction assumed to be a
random Gausian function. This problem is relevant for a variety
 of important physical
situations, ranging from electromagnetic waves propagating through
the interstellar space or the atmosphere, to electron transport in
disordered conductors. The  wave density, $I({\bf r})= |\psi({\bf
r})|^2$, exhibits sample specific random fluctuations (speckles) due
to interference of waves traveling along different paths. The
statistical properties of speckles have been discussed  in the past.
The problem can be characterized by several characteristic lengths:
the propagation distance, $Z$, the elastic mean free path, $\ell$
and  the transport length $\ell_{tr}\sim \ell/\theta_{0}$, which is
the typical distance for backscattering. Here $\theta_{0}\sim k\xi$
is the typical scattering angle on the distance $\ell$, and $\xi$ is
the correlation length of $n({\bf r})$. In the diffusive regime $Z
\gg \ell_{tr}$ the problem has been studied in Refs.
\cite{ZyuzinSpivak,KaneLee,ZyuzinSpivakRev}. The limit of "directed
waves" $\ell_{tr}\gg Z\gg \ell$ has been studied in many papers, see
for example ~\cite{Tatatarski,Kravtsov,Prokhorov,Dashen} and
references therein. In the latter case the wave experiences many
small angle scattering events, but the total change of its
propagation angle $\theta$ remains small.

\begin{figure}[ptb]
\includegraphics[width=8.5cm]{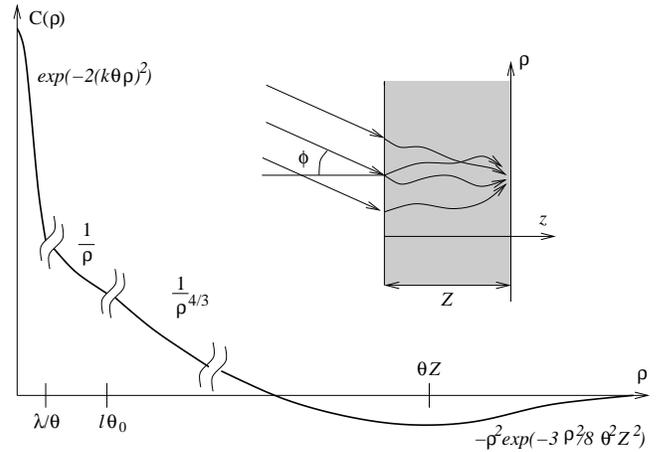}
\caption{The asymptotic behavior of the density correlation function
$C(\rho)=\langle \delta I(\mathbf{\rho}) \delta I(0) \rangle$.
$\lambda$ is the light wavelength, $\ell$ is the elastic mean free
path, $Z$ is the slab width, and $\theta_0$ and $\theta$ are the
typical scattering angle of a ray traveling a distance $\ell$ and
$Z$ respectively.} \label{fig:main}
\end{figure}

In this Letter we develop a general method for calculating speckle
correlations, which enables us to treat both diffusive and directed
wave cases on equal footing. It is similar, in some of its aspects,
to the Langevin scheme for the description of classical fluctuations
\cite{LandauLifshitz,ShulmanKogan,KoganBook}. We shall demonstrate
the method for the case $\ell_{tr}\gg Z\gg \ell$ by calculating the
speckle correlations and their sensitivity to various perturbations,
such as a change in the  frequency  of the wave, its incidence angle
and a change of the refraction coefficient. In many aspects, our
results differ from those obtained in the previous
studies~\cite{Tatatarski,Kravtsov,Prokhorov,Dashen}. Among the
differences are the slow power law decay of the density correlator
as a function of coordinates and its change of sign, see
Fig.~\ref{fig:main}.

The central object of our approach is the ray distribution function
$f({\bf r},{\bf s})$
\begin{equation}
f({\bf r},{\bf s})\!= \!\int\!\! \frac{p^{2}dp}{2\pi^2} \int\! \!d
\mathbf{r}' \psi\left({\bf r}\!-\!\frac{{\bf r'}}{2}\right)
\psi^*\left( {\bf r} \!+\!\frac{{\bf r'}}{2} \right)e^{i p{\bf
s}\cdot {\bf r'}},
\end{equation}
which is the probability of finding a ray at point ${\bf r}$
pointing in the direction specified by the unit vector ${\bf s}$, in
particular $I({\bf r})=\int d{\bf s} f({\bf r,s})$. The average
distribution function $\langle f({\bf r,s}) \rangle$ satisfies the
Boltzmann kinetic equation,
\begin{eqnarray}\label{eq:kinetic}
{\bf s}\cdot \frac{\partial \langle f({\bf r,s})\rangle}{\partial
{\bf r}}\!&\!=\!& \!I_{st}\{\langle f({\bf r,s})\rangle\}, \\
 I_{st}\{\langle f\rangle\}\!&\!=\!&\!\!\int
\!d^{2}\tilde{s} W({\bf s\!-\!\tilde{s}}) \left[ \langle f({\bf
r},\tilde{\bf s})\rangle -\langle f({\bf r},{\bf s}) \rangle
\right].
 \label{eq:collision}
\end{eqnarray}
Here $\langle \ldots \rangle$ denotes averaging over the random
realizations of $n({\bf r})$, the integral over the directions is
normalized to unity, $\int d^2 s =1$, and $W({\bf \delta
s})=\frac{k^4}{\pi} \int d^{3}r g({\bf r}) e^{i k \delta {\bf s}
\cdot {\bf r}}$ is the probability, per unit length of propagation,
for changing the ray direction by $\delta {\bf s}$, and $ g({\bf
r}-{\bf r'})= \langle n({\bf r})n({\bf r'})\rangle - \langle n({\bf
r})\rangle \langle n({\bf r'})\rangle $ is the disorder correlation
function.
\begin{figure}[ptb]
\includegraphics[width=8cm]{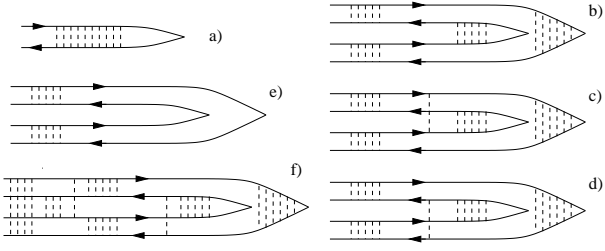}
\caption{Diagrams describing wave density correlations. The solid
lines denote the disorder-averaged Green functions of
Eq.~(\ref{eq:wave}) and the dashed lines denote the disorder
correlator, $4 k^4 \langle \delta n(\mathbf{r})\delta
n(\mathbf{r}')\rangle$. } \label{fig:diagrams}
\end{figure}

Correlations of the fluctuations, $\delta f=f-\langle f\rangle$, may
be evaluated using the Langevin-type equation,
\begin{equation}
{\bf s}\cdot \frac{\partial \delta f}{\partial {\bf r}}-
I_{st}\{\delta f \}= {\cal L}({\bf r}, {\bf s}), \label{eq:langevin}
\end{equation}
where ${\cal L}$ is a random Langevin source, with a zero mean and
the variance given by,
\begin{widetext}
\begin{equation}
\langle {\cal L}({\bf r},{\bf s}) {\cal L}({\bf r'}, {\bf
s'})\rangle \!=\! \frac{2\pi }{k^2}\delta({\bf r}\!-{\bf r'})
 \left[ \delta({\bf s}\!-\!{\bf s'})\langle
f({\bf r,s})\rangle \int d^{2} s_{1} W({\bf s}\!-\! {\bf s}_{1})
\langle f ({\bf r}, {\bf s}_{1})\rangle - \langle f({\bf
r,s})\rangle W({\bf s}\!-\!{\bf s'}) \langle f ({\bf r},{\bf
s'})\rangle \right].
 \label{eq:langevinsources}
\end{equation}
\end{widetext}
One can prove Eqs.(\ref{eq:kinetic}-\ref{eq:langevinsources}) using
the standard impurity diagram technique~\cite{Abrikosov}. Equations
(\ref{eq:kinetic},\ref{eq:collision}) are obtained by summing the
ladder diagrams in Fig.~\ref{fig:diagrams} a), whereas
Eqs.~(\ref{eq:langevin},\ref{eq:langevinsources}) follow from
diagrams b)-d) in Fig.~\ref{fig:diagrams}.

Equations (\ref{eq:kinetic}-\ref{eq:langevinsources}) are valid when
the mean free path is sufficiently large, $\ell = (\int d s^{2}
W({\bf s}))^{-1}\!\gg\! \xi^{2}/\lambda$, , and
$|\mathbf{r}-\mathbf{r}'|\!\gg\! \lambda$. Here ${\bf r}$ and ${\bf
r'}$ are observation points, $\xi=[\int d^3r r^2 g(r) /3 \int d^3r
g(r)]^{1/2}>\lambda$ is the disorder correlation length, and
$\lambda=2\pi/k$  is the wave length. For $|{\bf r}-{\bf r'}|\leq
\lambda$ Eqs.~(\ref{eq:kinetic}-\ref{eq:langevinsources}) are not
valid and to evaluate the correlation function  one has to calculate
the diagram shown in Fig.~\ref{fig:diagrams} e).

In the regime of small angle scattering, $\theta_{0}\ll 1$, and when
$|{\bf r}-{\bf r}'| \gg l$ the rays undergo diffusion in the space
of directions, ${\bf s}$. In this case
 Eqs.~(\ref{eq:kinetic}-\ref{eq:langevinsources}) reduce to a
set of angular diffusion equations, \label{eq:langevin_angle}
\begin{eqnarray}\label{eq:langevin_angle_1}
{\bf s} \cdot \frac{\partial \langle f({\bf r,s}) \rangle}{\partial
{\bf r}}&=& D_\theta \nabla_s^2
 \langle f({\bf r,s}) \rangle, \\
 \label{eq:langevin_angle_2}
{\bf s} \cdot \frac{\partial \delta f({\bf r,s})}{\partial {\bf r}}
&=&\nabla_s \left[
 D_\theta  \nabla_s
 \delta f({\bf r,s}) -{\bf j}_{L}({\bf r},{\bf s})\right], \\
 \label{eq:langevin_angle_3}
\langle j^L_{\alpha} ({\bf r},{\bf s}) j^L_{\beta}(\tilde{\bf
r},\tilde{\bf s})\rangle &=&\frac{2 \pi D_{\theta}\langle f\rangle
^{2}}{k^{2}}\delta_{\alpha\beta} \delta ({\bf s}\!-\!\tilde{\bf s})
\delta ({\bf r}\!-\! \tilde{\bf r}).
\end{eqnarray}
Here ${\bf j}^{L}({\bf r},{\bf s})$ are the
Langevin current sources,
$D_\theta=\frac{1}{2}\ell_{tr}^{-1}=\frac{1}{2}\int d^2
s'(1-\mathbf{s}\cdot \mathbf{s}')W(\mathbf{s}-\mathbf{s}')$
is the diffusion constant in the space of angles ${\bf s}$, and $
\nabla_{s}= \hat{\theta} \frac{\partial}{\partial \theta}+
\frac{\hat{\phi}}{\sin(\theta)} \frac{\partial}{\partial \phi} $ is
the gradient operator, with $\hat{\phi}= (-\sin \phi,\cos \phi,0)$,
and $\hat{\theta}=(\cos \phi \cos \theta, -\sin \phi\cos \theta
,-\sin \phi)$.

Further simplification emerges at larger spatial scales,  $|{\bf
r}-{\bf r'}|\gg \ell_{tr}$. In this case
Eqs.~(\ref{eq:kinetic}-\ref{eq:langevinsources}) can be reduced to
diffusion equations~\cite{ZyuzinSpivak,ZyuzinSpivakRev}, $\nabla^2
\langle I \rangle =0$, and $\nabla (- D \nabla \delta I +{\bf
J})=0$. Here $D=\ell_{tr}/3$  is the (real space) diffusion
constant, and correlation function for Langevin currents has the
form: $\langle J_\alpha({\bf r}) J_\beta({\bf r'})\rangle=
\frac{\lambda^2 \ell_{tr}}{6\pi}
 \langle I ({\bf r}) \rangle^{2} \delta_{\alpha \beta} \delta ({\bf r}-{\bf r'})$.
In this case, the correlation function $\langle \delta I({\bf r})
\delta I({\bf r'}) \rangle$ decays as $1/r$ and as $1/r^2$ for $r
\gg \ell_{tr}$ and $\lambda \ll r \ll \ell_{tr}$ respectively.

The scheme presented above  can be generalized to treat speckle
sensitivities to changes of external parameters, such as the
wavenumber, $\Delta k$, or smooth changes of the refractive index,
$\Delta n({\bf r})$. We characterize the sensitivities by the
correlation function $\langle \delta f(0)\delta f(\gamma) \rangle $,
where $\gamma=\Delta k+k \Delta n({\bf r})$. In this case the
Langevin source correlator is given by,
\begin{widetext}
\begin{equation}
\langle {\cal L}({\bf r},{\bf s};0) {\cal L}({\bf r'}, {\bf
s'};\gamma )\rangle \!=\! \frac{\pi}{k^2}\delta({\bf r}\!-{\bf r'})
\sum_{\nu=\pm} \left[ \delta({\bf s}\!-\!{\bf s'}) f_\nu({\bf r,s})
\int d^{2}{\bf s}_{1} W({\bf s}\!-\!{\bf s}_{1}) f_{-\nu}({\bf
r},{\bf s}_{1}) - f_\nu({\bf r},{\bf s}) W({\bf s}\!-\!{\bf s'})
f_{-\nu}({\bf r},{\bf s'}) \right], \label{eq:langevin-sensitivity}
\end{equation}
\end{widetext}
where $f_\pm ({\bf r,s})$ satisfies the equation,
\begin{equation}
{\bf s} \cdot \frac{\partial  f_\pm({\bf r,s})} {\partial {\bf r}}-
I_{st}\{f_{\pm }({\bf r,s})\}=\pm i \gamma  f_\pm({\bf r,s})
.\label{eq:kinetic-senstitivity}
\end{equation}

The method based on
Eqs.~(\ref{eq:kinetic}-\ref{eq:kinetic-senstitivity}) is similar to
the Langevin approach describing classical time and space
fluctuations of a single particle distribution function
\cite{LandauLifshitz,ShulmanKogan,KoganBook}. The fundamental
difference between our problem and that of classical fluctuations
manifests itself in the form of the correlation function of the
Langevin sources. In the classical problem the Langevin sources are
$\delta$-correlated both in time and space and their variance is
proportional to the average distribution function $\langle
f\rangle$. In contrast, in
Eqs.~(\ref{eq:langevinsources},\ref{eq:langevin-sensitivity}) the
Langevin source variance is quadratic in $\langle f\rangle$ and
$\delta$-correlated only in space.

To illustrate the use of
Eqs.~(\ref{eq:kinetic}-\ref{eq:kinetic-senstitivity}) we consider
the case when a  wave of intensity $I_0$ is incident on a disordered
slab of thickness $Z$, such that $\ell \ll Z\ll \ell_{tr}$, as shown
in the inset of Fig.~\ref{fig:main}. The results presented below are
calculated to leading order in the small total scattering angle,
$\theta^2\equiv D_\theta Z$. In this case the correlation function
${\cal C}({\bf r-r'}) = \langle \delta I({\bf r}) \delta I({\bf r}')
\rangle$ is strongly anisotropic (here $\delta I({\bf r})= I({\bf
r})-\langle I({\bf r}) \rangle$). Therefore below we shall use the
notation: ${\bf r}=(z,\vec{\rho})$, where $\vec{\rho}$ denotes a
two-component vector in the plane perpendicular to the $z$-axis and
$z$ denotes the distance between observation points along the $z$
axis. When $\rho=0$, i.e the observation points are located along
the $z$ axis, the correlation function for  $ z\ll Z$ is,
\begin{equation}
{\cal C}(z)=\frac{I_0^2}{4k^2\theta^4 z^2}. \label{eq:C}
\end{equation}
Equation (\ref{eq:C}) matches the results for the diffusive case
\cite{ZyuzinSpivak,ZyuzinSpivakRev}, $Z\gg \ell_{tr}$ when $\theta$
is of order unity.

When $z<\rho/\theta$,  i.e. the observation points are located
essentially on a plane perpendicular to the $z$ axis, a general
formula for ${\cal C}(\mathbf{\rho})$, can be derived from
Eqs.~(\ref{eq:langevin_angle_1}-\ref{eq:langevin_angle_3}),
\begin{eqnarray}
{\cal C}(\rho)&=&\frac{I_0^2}{4 D_\theta k^2}
\int_0^{Z-\ell}\!\frac{d\zeta}{\zeta-Z}\int_0^\infty \!\! dq q J_0(q
\rho)
\nonumber \\
&&\times \frac{d}{d\zeta} \exp \left[-\frac{2}\ell \int_0^\zeta
d\eta \left\{ 1- \tilde{g}\left(\frac{q}{k}\eta\right)\right\}
\right], \label{eq:main}
\end{eqnarray}
where $\tilde{g}(\rho)=\int dz g(\sqrt{\rho^{2}+z^{2}})/\int dz
g(z)$. The integral in Eq.~(\ref{eq:main}) contains a term
proportional to a $\delta$-function, $\frac{\pi I_0^2}{2 D_\theta
 k^2 Z}\delta(\vec{\rho})$. This term represents the rapidly
decaying (at $\rho \sim \lambda/\theta$) part of the correlator and
corresponds to the part of diagram Fig.~\ref{fig:diagrams} b)
without the impurity ladders after the Hikami box, see diagram e).
The $\delta$-function term results from the semiclassical
approximation employed in the derivation of
Eqs.~(\ref{eq:kinetic}-\ref{eq:kinetic-senstitivity}), which limits
the spacial resolution to $\delta \rho \gg \lambda$. In order to
resolve the spacial structure of the short distance part of the
correlator diagram e) needs to be evaluated more accurately. This
gives the following asymptotic behavior,
\begin{equation}
\frac{{\cal C}(\rho)}{I_0^2}\approx \left \{
\begin{array}{ll} e^{-2(k \theta \rho )^2} &
\textrm{if~~~ $\rho \sim\alpha
 \lambda /\theta$}, \\
\frac{b_1}{k^2 \theta^2\theta_0 \rho} &   \textrm{if~~~ $\alpha
\lambda / \theta \ll \rho\ll \ell \theta_0$}, \\
 \frac{ b_2 D_\theta^{2/3}}{k^{2}
\theta^4 \rho^{4/3}} & \textrm{if~~~ $ \ell \theta_0 \ll \rho \ll
\theta Z $} ,\\
\frac{-b_3 \rho^2}{k^2 \theta^6 Z^4}e^{-\frac{3 \rho^{2}}{8
  \theta^2 Z^2}} & \textrm{if~~~ $\theta Z \ll \rho \ll
 \frac{Z\theta^2}{\theta_0}$},
\end{array} \right.
\label{eq:asymptotic}
\end{equation}
where $\alpha^2=\log(k \ell \theta^3/\theta_0)$, $b_1=\int_0^\infty
dx \tilde{g}(x)$ is a constant of order unity, $b_2=3^{1/3}\Gamma
(5/3)/8\approx 0.163$, and $b_3=27/128\approx 0.21$. The tail of the
correlation function (the regime $\rho>Z\theta^2/\theta_0$) is also
described by Eq.~(\ref{eq:main}) and depends on the precise form of
the disorder correlation $g(r)$, since this limit is dominated by
rare scattering events. The qualitative form of the function ${\cal
C}(\rho)$ is shown in Fig.~\ref{fig:main}.

Let us consider now the statistics of density, integrated over a
disk of radius $R$, $P=\int_{\rho \leq R} d^2 \rho
I(\mathbf{\rho})$. Using Eqs.~(\ref{eq:main}, \ref{eq:asymptotic})
we get,
\begin{equation}
\frac{\langle(\delta P)^2\rangle }{I_0^2 \pi^2 R^4}\approx \left \{
\begin{array}{ll}
\frac{\pi}{2k^2 \theta^2}+\frac{b'_1 (D_\theta R)}{k^2
\theta^4\theta_0} ,&  \frac{\alpha
\lambda}{ \theta} \! \ll \! R \! \ll \! \ell \theta_0 ,\\
 \frac{\pi}{2k^2 \theta^2}+\frac{b'_2 (D_\theta R)^{2/3}}{k^2
 \theta^4},
 & \,\ell \theta_0 \! \ll \! R \!\ll\!
\theta Z , \\
b'_3 \frac{Z}{\theta R}, & \theta Z \! \ll \! R \! \ll \!
 \frac{Z \theta^2}{\theta_0},
\end{array} \right.
\label{eq:power-asymptotic}
\end{equation}
where $b'_1=2b_1\pi/3$, $b'_2=3^{4/3}\Gamma(5/6) \pi/2^{11/3}
\Gamma(7/6)$, and $b'_3=\sqrt{3/2\pi}$.

Consider now the sensitivity of the integrated density $P({\omega})$
to a change in the wave frequency $\Delta \omega=c\Delta k$, where
$c$ is the speed of the wave.  It can be characterized by the
experimentally accessible  quantity,
\begin{equation}
 \frac{\langle \left( P(\omega\!+\!\Delta \omega)\!-\!P
(\omega)\right)^2\rangle }{ \langle \left( \delta P \right)^2
\rangle}\approx \left(\frac{\Delta \omega}{\omega^*}\right)^2,
\end{equation}
where
\begin{equation}
\omega^{*}= \sqrt{\frac{15}{2}} \frac{c}{\theta^2 Z}.
 \label{eq:omega*}
\end{equation}
A qualitative explanation of the scale $\omega^*$ is similar to that
given for the sensitivity of the conductance fluctuations
\cite{LeeStone,AltshulerSpivak}. Let us estimate the characteristic
change in the phase of a typical orbit due to the frequency change
$\Delta \omega$ : The typical length spread of the orbits is of
order $\theta^2 Z$. Therefore the phase difference is $\Delta k Z
\theta^2$ where $\Delta k=\Delta \omega/c$ is the change in the
wavenumber. Thus a complete change of the speckle pattern occurs
when the phase, $\Delta \omega Z \theta^2/c$ is of order one, namely
$\Delta \omega \sim c/Z\theta^2$, in agreement with
Ref.~\cite{Dashen}.

As another application of our scheme let us consider the sensitivity
of speckles to a change in the incidence angle, $\phi$, of the wave,
see the inset in Fig.~\ref{fig:main} (thus the incident wave
function now has a form  $\psi=\sqrt{I_0} \exp[i k z\cos\phi + ik
\rho\sin\phi ]$). One may characterize this sensitivity by the
correlation function,
\begin{equation}
 \frac{\langle\delta P(\phi)\delta P
(0)\rangle }{ \langle \left( \delta P \right)^2 \rangle}\approx
e^{-\frac{2}{3}\frac{\phi^2}{\phi^{*2}}} +\frac{\beta \left(
e^{-\frac{1}{8}\frac{\phi^2}{\theta^2}}-
e^{-\frac{2}{3}\frac{\phi^2}{\phi^{*2}}} \right)}{ k^2 \ell^4
    D_\theta^2},
\label{eq:phi-sensitivity}
\end{equation}
where $\phi^{*}= (\theta k Z)^{-1}$, $\beta$ is a factor of order
unity, and it is assumed that $R\ll \theta Z$. The first term of
this equation follows from formula (\ref{eq:langevin}) and the
correlator (\ref{eq:langevin-sensitivity}), where $f_\pm$ satisfies
equation (\ref{eq:kinetic-senstitivity}) with initial conditions
\begin{equation}
f_\pm(\vec{\rho}, z=0)=  I_0 e^{\pm i k {\bf s}_\perp
  \vec{\rho}}\delta ({\bf s}-{\bf s}_0),
\end{equation}
where ${\bf s}_0=(\cos \phi,\mathbf{s}_\perp)\approx
(1,\mathbf{s}_\perp)$, with $|\mathbf{s}_\perp|=\sin \phi\approx
\phi$, assuming $\phi \ll 1$. The second term in the right hand side
of Eq.~(\ref{eq:phi-sensitivity}) is computed from diagrams of the
type shown in Fig.~\ref{fig:diagrams} d), containing two Hikami
boxes. It represents a small correction in the parameter
$\xi/\ell\theta_0$, however, this term becomes the dominant
contribution when $\phi \gg \phi^*$.

On a more general level
Eqs.~(\ref{eq:kinetic}-\ref{eq:kinetic-senstitivity}) provide a
framework for calculation of time correlation of speckle patterns in
the case where the refraction index changes in time, provided these
changes are much slower than the time of propagation of the wave to
the observation point.

Our results also may be easily extended to cases with light
polarization, optically active media, Faraday effect, and coherent
short wave pulses as long as their duration is longer than
$\tau=\ell/c$. These issues are left for future studies.

The  problem considered  here is similar to the problem of universal
conductance  fluctuations in metallic samples
\cite{LeeStone,Altshuler}, which are also of interference nature.
Therefore we would like to discuss the relation between the two
problems. In the single particle approximation the conductance of a
metallic sample,  $G\sim \int d\phi \, T(\phi)$ can be expressed in
terms of the electron transmission probability through the sample,
$T(\phi)\propto \int d\mathbf{\rho} \, d^2 s  \,({\bf z} \cdot {\bf
s}) f(\mathbf{\rho},{\bf s})$, integrated over the incidence angle
of the incoming wave, $\phi$. Thus the variance of the conductance
fluctuations, $\delta G$ is proportional to a double integral of the
correlation function $\langle \delta T(\phi)\delta T(\phi')\rangle$.
In principle, the latter can be calculated using
Eqs.~(\ref{eq:kinetic}-\ref{eq:kinetic-senstitivity}), or,
equivalently, by calculating diagrams shown in
Fig.~\ref{fig:diagrams} b)-d). However, as we explain below, this
does not account for the conductance fluctuations. The latter arise
from diagrams of the from shown in Fig.~\ref{fig:diagrams} f).

In the limit of directed waves, $Z\ll \ell_{tr}$, there is no
backscattering. Therefore the transmission probability does not
fluctuate, $\delta T(\phi)\sim \delta G=0$. Thus to compare the two
problems we have to consider the diffusive case, $Z\gg l_{tr}$,
where $ \delta G\sim e^{2}/\hbar$. The correlation function $\langle
\delta T(\phi)\delta T(\phi')\rangle$, in the diffusive regime,
still has a structure similar to the correlation function given by
Eq.~(\ref{eq:phi-sensitivity})~\cite{ZyuzinSpivakRev}. Namely, it
contains two contributions. The first contribution comes from
diagrams \ref{fig:diagrams} b)-d), and describes relatively strong
fluctuations of the transmission coefficient. However, it is very
sensitive to the change of $\phi$, and after the integration over
$\phi$ gives a small contribution to $\delta G^2$. The second
contribution  originates from diagrams of the type
Fig.~\ref{fig:diagrams} f), and is analogous to the second term in
Eq.~(\ref{eq:phi-sensitivity}). Although its amplitude is smaller
than that of the first term, it is insensitive to the change of
$\phi$, and after the integration over $\phi$ yields the dominant
contribution to $\delta G^2$.  Thus, conductance fluctuations are
not described by
Eqs.~(\ref{eq:kinetic}-\ref{eq:kinetic-senstitivity}), and should be
calculated from the diagrams of the type shown in
Fig.~\ref{fig:diagrams} f) (see the corresponding discussion in Ref.
\cite{ZyuzinSpivakRev}).

Finally we would like to mention that the results presented above
substantially differ from those known in the literature (see for
example Refs.~\cite{Tatatarski,Kravtsov,Prokhorov,Dashen}). First,
the correlation function (\ref{eq:main}) exhibits a universal long
range power law behavior in a wide range of values of $\rho$. The
only  non-universal regimes are at the tail, $\rho\gg Z
\theta^2/\theta_{0}$ , and the short distance region, $\rho \sim
\xi$. In contrast, in the results presented in
Refs.~\cite{Tatatarski,Kravtsov,Prokhorov,Dashen}  ${\cal C}(\rho)$
depends on  the detailed form of $g(r)$, and usually decays
exponentially at $\rho>\xi$. Second, in contrast with previous
results, ${\cal C}(\rho)$ changes its sign as a function of $\rho$,
which is a consequence of the current conservation. This
conservation law also implies, that the fluctuations of the
integrated intensity over disks of radius $R>Z\theta$ is
proportional to $R$, see Eq.~(\ref{eq:power-asymptotic}), rather
than $R^2$, as would follow from
Refs.~\cite{Tatatarski,Kravtsov,Prokhorov,Dashen}.

Useful discussions with A. Zyuzin are acknowledged. This work has
been supported by the Packard Foundation, by the NSF under Contracts
No. DMR-0228104, and  by the Israel Science Foundation (ISF) funded
by the Israeli Academy of Science and Humanities, and by the
USA-Israel Binational Science Foundation (BSF).

\end{document}